# Enhanced interferometric resolution via N-fold intensity-product measurements without sacrificing phase sensitivity

S. Kim[1], J. Stöhr[2], and Byoung S. Ham[1,*]

[1]Department of Electrical Engineering and Computer Science, Gwangju Institute of Science and Technology, Gwangju 61005, South Korea
[2]SLAC National Accelerator Laboratory and Department of Photon Science, Stanford University, CA 94305, USA
(June 06, 2025; *bham@gist.ac.kr)

**Abstract**
The Fisher information theory sets a fundamental bound on the minimum measurement error achievable from independent and identically distributed (*i.i.d.*) measurement events. The assumption of identical and independent distribution often implies a Gaussian distribution, as seen in classical scenarios like coin tossing and an optical system exhibiting Poisson statistics. In an interferometric optical sensing platform, this translates to a fundamental limit in phase sensitivity – known as the shot-noise limit (SNL)– which cannot be surpassed without employing quantum techniques. Here, we, for the first time to the best of our knowledge, experimentally demonstrate a SNL-like feature on resolution of an unknown signal when intensity-product measurement technique is applied to N-divided MZI output subfields. Given the Poisson-distributed photon statistics, the N-divided subfields ensure the *i.i.d.* condition required by Fisher information theory. Thus, the N-fold intensity-product technique holds promise for enhancing the precision of conventional optical sensing platforms such as a fiber-optic gyroscope and wavelength meter, while preserving the original phase sensitivity of the output field.

**Introduction**
Optical sensing and metrology have been one of the most important research areas in modern science and technologies [1-13]. In optical sensing, high-resolution spectroscopy has been the key interest in physics [1-3], chemistry [4,5], biology [6,7], medicine [7], semiconductor industries [8,9], and military services [10,11]. The phase sensitivity of an unknown optical signal is given by Fisher information theory leading to the lower bound of the shot-noise limit (SNL) originated in photon statistics of Poisson distribution [12]. According to the Fisher information theory, the noise of independent and identically distributed (*i.i.d.*) measurement events increases as $\sqrt{N}$, when the signal does as N, resulting in $\sqrt{N}$ gain in phase sensitivity. Thus, an easy way to get high sensitivity is to use a high power laser, as in LIGO interferometry for the gravitational wave detection [13]. While the Heisenberg limit in quantum sensing offers a $\sqrt{N}$ advantage over SNL, the achievable photon number N in N00N states is severely constrained, typically remaining well below 100. Consequently, the theoretical $\sqrt{N}$ gain is practically outweighed by classical sensors utilizing laser light with photon numbers many orders of magnitude higher. Although N00N-based quantum sensing provides N-fold enhancement in resolution, as demonstrated by the photonic de Broglie wavelength [14], classical techniques employing optical cavities or interferometers have already achieved resolutions on the order of $10^{-7}\lambda$ in wavelength meters and $10^{-7}\Omega_E$ in ring-laser gyroscopes [16]—levels unattainable with current N00N-state implementations.

In optical sensing, resolution is typically defined by the Rayleigh criterion, which states that two sources are resolvable when the principal maximum of one diffraction pattern coincides with the first minimum of the other. In interferometer-based systems such as MZI and optical cavity, resolution is more practically characterized by the full-width at half maximum (FWHM) of the interference fringes [17]. Notably, the fringe resolution remains independent of the signal's intensity unless it is in a single-photon regime as in quantum sensing. Importantly, resolution should be distinguished from phase sensitivity, where phase sensitivity involves variance to the slope of the signal. Thus, the minimum phase sensitivity often arises in regions of low signals, which contrasts with conventional applications, such as laser locking [18], radar [19], lidar [20], biosensors [21], and ring-laser gyroscope [22], where best sensitivity is sought in regions of steep signal change.

Unlike quantum sensing approaches that utilize higher-order intensity correlation via coincidence detection [14,15], most classical sensing techniques are based on first-order intensity correlation [1-12,16,18-22]. Recently, intensity-product measurements have been introduced into classical sensing frameworks to surpass the diffraction limit [23,24]. Here, we report, to the best of our knowledge for the first time, the experimental demonstration of MZI fringe-resolution enhancement using higher-order intensity products. This resolution improvement exhibits a SNL scaling behavior in terms of fringe FWHM. However, within one phase period, the observed $\sqrt{N}$ resolution gain at a given phase point is offset by a $\sqrt{N}$ resolution loss at another across the phase domain. We analyze this $\sqrt{N}$-enhanced resolution and examine its relation to phase sensitivity through the framework of Fisher information. While this method does not deteriorate SNL, it offers a viable strategy for resolution enhancement in interferometer-based optical sensors, such as a fiber-optic gyroscopes and wavelength meters that operate with high-power light sources.

**Results**

Figure 1 shows the schematic of the intensity product-enhanced sensing metrology in MZI. For this, one MZI output port is evenly divided into four ports for up to fourth-order intensity-product measurements, where measurement scheme of the intensity product between subdivided ports is for either a single-photon or continuous-wave (CW) regime. Importantly, the intensity-product order N is scalable up to the photon number of the input light [24]. This practical benefit enables performance that may even outperform the $\sqrt{N}$ scaling characteristic of quantum enhanced sensing. For measurements, the usual coincidence-detection method in quantum metrology is adopted for a single-photon regime [14,15]. For a continuous-wave (CW) regime, the single-photon detectors are simply replaced by photodetectors. Regarding the four-split MZI output ports, global phases generated by inserted BSs [25] and added path lengths to individual detectors from the MZI do not affect their intensities due to Born's rule stating that measurement is the absolute square of its amplitude. Due to Poisson statistics of coherent lights, thus, all divided subfields satisfy the *i.i.d.* condition of random variables.

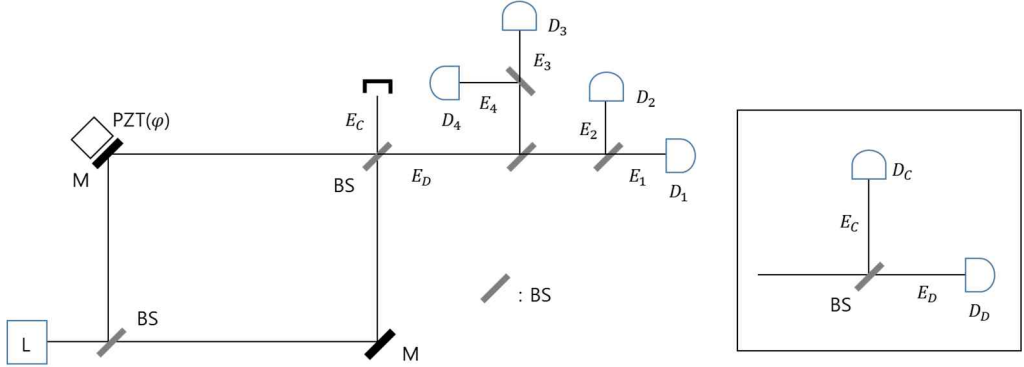

**Fig. 1.** Schematic of higher-order intensity correlation. L: laser, BS: 50/50 nonpolarizing beam splitter, $D_j$: photodetector (or single photon counting module), M: mirror, PZT: piezo-electric transducer. (Inset) Cross-product.

For the coherence solution of the intensity products between divided subfields in Fig. 1, both MZI output intensities are firstly derived:

$$I_C = I_0(1 - \cos\varphi), \qquad (1)$$

$$I_D = I_0(1 + \cos\varphi), \qquad (2)$$

where $I_C = E_C E_C^*$ and $I_D = E_D E_D^*$. Here, the input intensity is set for $2I_0$ for simplicity. From Eqs. (1) and (2), it is clear that the MZI has a deterministic coherence feature of interference fringes regardless of input intensity $2I_0$. This makes the fringe resolution of conventional interferometry is independent of photon numbers or input intensity. Including an arbitrary global phase $e^{i\zeta_j}$ at each detector, the first-order intensity correlation of all divided subfields is identical:

$$I_j = \frac{I_0}{N}(1+\cos\varphi). \qquad (3)$$

As mentioned above, the biggest N is limited by the photon number of the input light $2I_0$: In a CW laser light at 1 μW, the photon number counted for one second is in the order of $10^{11}$. For the single-photon regime, N coincident photons are post-selected by a coincidence counting unit, where a few percent error is inevitable due to the Poisson statistics. In Fig. 1, thus, the generalized Nth-order intensity correlation can be represented by:

$$R_D^{(N)} = \left(\frac{I_0}{N}\right)^N (1+\cos\varphi)^N. \qquad (4)$$

Here, the intensity product term is conditioned for $\left(\frac{I_0}{N}\right)^N = 1$ for the single-photon regime and $\left(\frac{I_0}{N}\right)^N \gg 1$ for the CW regime. Thus, the *i.i.d.* joint product in Eq. (4) gives a great sensing benefit over the first-order intensity correlation with an input of a commercially available laser (see Discussion).

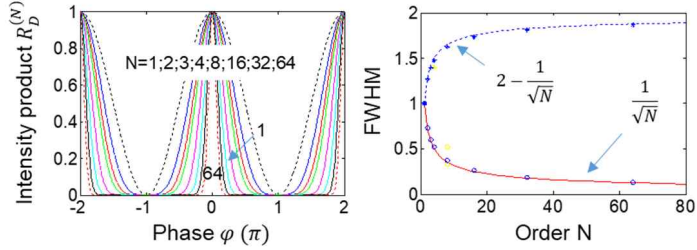

**Fig. 2.** Numerical calculations of the normalized intensity product. (left panel) For N=1 (dotted), N=2 (blue), N=3 (red), N=4 (green), N=8 (magenta), N=16 (cyan), N=32 (black), N=64 (red-dotted). (right panel) Red (dotted) curve is for $1/\sqrt{N}\left(2 - 1/\sqrt{N}\right)$. Circles: FWHM data from the left panel at $\varphi = 0$. Stars: FWHM data from the left panel at $\varphi = \pm\pi$.

Figure 2 presents numerical results for the Nth-order intensity correlations calculated using Eq. (4), where N denotes the number of divided output ports. All intensity products are normalized for comparison purposes (see the left panel). As shown in the right panel, FWHM of the Nth-order intensity correlations near $\varphi = 0$ scales as $1/\sqrt{N}$, demonstrating resolution enhancement (see circles on the red curve). In contrast, the resolution deteriorates near $\varphi = \pm\pi$, as indicated by stars on the dotted curve. The data points marked by circles and stars represent the FWHMs extracted from the left panel. These results confirm that the enhanced resolution at near $\varphi = 0$ exhibits SNL-like scaling, albeit at the expense of reduced resolution at $\varphi = \pm\pi$ – a tradeoff not reported yet. In other words, Fig. 2 demonstrates that the higher-order intensity-product measurement surpasses the diffraction limited resolution with N=1, in the region of constructive inference. Notably, this technique is entirely classical and can be readily implemented in conventional interferometer-based sensing systems. Under Poisson statistics, the N-divided output subfields are treated as *i.i.d.*, justifying the statistical ensemble used in the analysis.

Figure 3 shows the experimental demonstration of resolution enhancement via the intensity-product measurements, as discussed in Fig. 2 and derived from Eq. (4). Results are shown for both the single-photon regime (Fig. 3(a)) and the CW regime (Fig. 3(b)) with corresponding numerical calculations for N=1,2,3,4 provided in Fig. 3(c). In Figs. 3(a) and (b), measured ratio of FWHMs between the Nth-order intensity products and the first-order (standard MZI, N=1) confirms the SNL-like feature of $\pi/\sqrt{N}$ near $\varphi \sim 0$, consistent with the predictions in Fig. 2. Conversely, near $\varphi \sim \pm \pi$, the FWHM ratio increases approximately as $\pi(2 - 1/\sqrt{N})$, reflecting a degradation in resolution that compensates for the enhancement at $\varphi \sim 0$. This tradeoff results in no net gain in global resolution over the full phase period. The observed local resolution enhancement at $\varphi \sim 0$ has not, to the best of our knowledge, been reported and is not anticipated by classical Fisher information. Moreover, the experimental results in Figs. 3(a) and (b) demonstrate that the FWHM of the normalized Nth-order intensity product, $R_D^{(N)}$, is independent of input power for a fixed order N, with fringe visibility remaining nearly ideal. Each data points in Figs. 3(a) and (b) represents an average over 30 samples with a 0.1 second integration time. The scanning mode used in Fig. 3(a) is consistent with that employed in commercially available wavelength meters to suppress noises of an interferometer, which would otherwise be dominated by Poisson fluctuation in the single-photon regime [26].

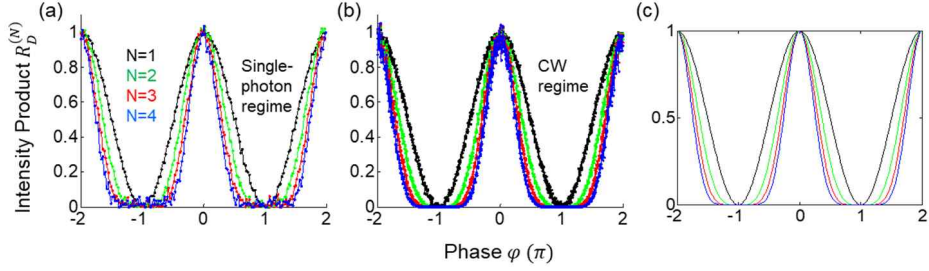

**Fig. 3.** Experimental demonstrations of higher-order intensity products. (a) Single-photon regime. (b) CW regime. (c) Numerical calculations. N = 1(black); 2(green); 3(red); 4(blue).

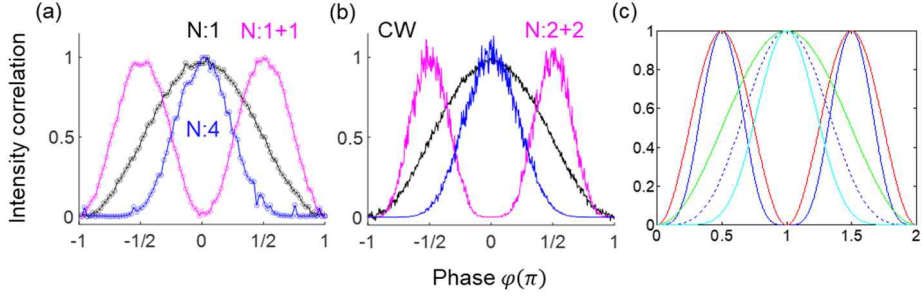

**Fig. 4.** Joint-intensity products between two MZI outputs in Fig. 1. (a) Single photon regime. (b) CW regime, (c) Numerical data. Black: N=1 for $I_D$. Blue: N=4 or $\prod_{j=1}^{4} I_j$. Magenta: N=1+1 or 2+2 for $I_C I_D$ (see the inset of Fig. 1). In (c), Green: N=1, Dotted: N=2, Cyan: N=4, Magenta: N=1+1, Blue: N=2+2.

Figure 4 compares the correlated intensity-product signal (magenta) with the uncorrelated ones in Fig. 3. The magenta curve represents the fourth-order cross-intensity correlation $R_{CD}^{(4)}$ between the two outputs of MZI, as defined by Eqs. (1) and (2) (see the inset of Fig. 1): $R_{CD}^{(2)} = I_C I_D = I_0^2 \sin^2 \varphi$ (Fig. 4(a)); $R_{CD}^{(4)} = \left(\frac{I_0}{2}\right)^4 \sin^4 \varphi$ (Fig. 4(b)). For comparison, the black curve represents the standard first-order intensity signal $R_D^{(1)}$ for N=1, while blue curve corresponds to the uncorrelated fourth-order intensity product $R_D^{(4)}$ in Eq. (4). Figures 4(a) and (b) show experimental results for the single-photon and CW regimes, respectively, and Fig. 4(c) represents the corresponding numerical calculations. The magenta fringe pattern, corresponding to the correlated measurement, exhibits twice the number of fringes compared to the uncorrelated $R_D^{(N)}$ signals. This doubling arises from the out-of-phase relationship between the two MZI outputs, reflecting an anti-correlation between the measurement events at ports C and D. In the single-photon regime (Fig. 4(a)), this anti-correlation in $R_{CD}^{(2)}$ leads to a doubling of resolution. In the CW regime (Fig. 4(b)), the correlated fourth-order signal $R_{CD}^{(4)}$ yields a resolution enhancement factor of $2\sqrt{2}$. The additional factor of $\sqrt{2}$ arises from the second-order intensity product contributions within each uncorrelated output, consistent with the scaling behavior discussed in Fig. 2. A generalized formulation for Nth-order correlated intensity products involving N/2 anti-correlated pairs has been recently proposed for classically excited superresolution [27].

**Discussion: Fisher information**

The Fisher information $F(\varphi)$ quantifies the amount of information that an observable random variable $X$ carries about an unknown parameter $\varphi$, and is defined by $F(\varphi) = (\partial_\varphi E[X])^2/\text{Var}(X)$ [12]. In Fig. 1, the N-divided output field of MZI, denoted as $I_j\ (= \frac{I_D}{N})$, serves as the observable random variable about an unknown parameter $\varphi$ due to the coherent input field satisfying Poisson statistics. In Eq. (1), variance $\text{Var}[I_D] = \langle I_D \rangle$ is satisfied due to Poisson nature of coherent light. This variance is inherently $\varphi$-dependent, and the vacuum noise becomes negligible in the limit $I_0 \gg 1$. To interpret the physical understanding of the Fisher information, one must consider the maximum slope of the signal $I_D$, which occurs at $\varphi = \pm\pi/2$. From the analytical expression $F(\varphi) = I_0(1 - \cos\varphi)$, however, the Cramer-Rao lower bound (CRLB) for phase estimation is minimized at $\varphi = \pi$, yielding $\Delta\varphi_{CRLB}^{(1)} \geq 1/\sqrt{2I_0}$. Thus, the minimum achievable phase sensitivity for the standard MZI output field $I_D$ is $\Delta\varphi_D = 1/\sqrt{2I_0}$, which corresponds to SNL.

In the same analogy, the Fisher information $F_S^{(N)}(\varphi)$ can be calculated for the N-fold intensity product signal $R_D^{(N)} = \left(\frac{I_0}{N}\right)^N (1 + \cos\varphi)^N$, as described in Eq. (4):

$$F_S^{(N)}(\varphi) = \frac{N^2\left(\frac{I_0}{N}\right)^{2N}\sin^2\varphi(1+\cos\varphi)^{2N-2}}{N\left(\frac{I_0}{N}\right)^{2N-1}(1+\cos\varphi)^{2N-1}} = I_0(1 - \cos\varphi), \tag{5}$$

where $\frac{d\langle S \rangle}{d\varphi} = N\left(\frac{I_0}{N}\right)^N (-\sin\varphi)((1 + \cos\varphi))^{N-1}$, $\text{Var}(S) = E[S^2] - (E[S])^2 = \mu^N\{\prod_{j=1}^N(1 + \mu_j) - \mu^N\} = \mu^N\{(1 + \mu)^N - \mu^N\} \sim N\mu^{2N-1}$, $\mu = I_j(\varphi)$, and $S \equiv R_D^{(N)}$. For $\mu \gg 1$, $(1 + \mu)^N - \mu^N \sim N\mu^{N-1}$ is obtained. Consequently, the phase sensitivity $\Delta\varphi_S$ of the N-fold intensity product $R_D^{(N)}$ is given by:

$$\Delta\varphi_S = \sqrt{\frac{1}{I_0(1-\cos\varphi)}}. \tag{6}$$

Interestingly, $\Delta\varphi_S$ in Eq. (6) is N-independent. Since the phase sensitivity of the MZI output field $I_D(\varphi)$ is given by $\Delta\varphi_D = \sqrt{\frac{1}{I_0(1-\cos\varphi)}}$, the phase sensitivity of the N-fold intensity-product signal $R_D^{(N)}$ remains identical to that of the standard MZI output $I_D$. This indicates that the N-fold intensity product preserves the phase sensitivity without degradation. Thus, the enhanced resolution observed in Figs. 3 and 4 are the benefit of the N-fold intensity-product measurements.

For the intensity product between two MZI outputs (see the magenta curve in Fig.4(a)), defined by $Xc(\varphi) = I_C I_D = I_0^2 \sin^2\varphi$, the corresponding Fisher information is derived as follows:

$$F_{Xc}^{(2)}(\varphi) = 2I_0\cos^2\varphi, \tag{7}$$

where $\frac{d\langle Xc \rangle}{d\varphi} = 2I_0^2 \sin\varphi\cos\varphi$ and $\text{Var}[Xc(\varphi)] = I_0^2 \sin^2\varphi(2I_0 + 1)$. Here, $\text{Var}[CD] = (E[C])^2\text{Var}(D) + (E[D])^2\text{Var}(C) + \text{Var}(C)\text{Var}(D) = \mu_C^2\mu_D + \mu_D^2\mu_C + \mu_C\mu_D = \mu_C\mu_D(\mu_C + \mu_D + 1)$. Accordingly, the resulting phase sensitivity is given by:

$$\Delta\varphi_{Xc} = \frac{1}{\sqrt{2I_0}\cos\varphi}. \tag{8}$$

As a result, the minimum phase sensitivity of Eq. (8) is equal to Eq. (6), resulting in $\Delta\varphi_{Xc} = \Delta\varphi_D = \Delta\varphi_S$. Therefore, both the N-fold intensity product shown in Fig. 3 and the cross-intensity product in Fig. 4 achieve enhanced resolution without any loss in phase sensitivity compared to the standard MZI output for the same input power. Even for the steepest slopes, $\Delta\varphi_{Xc}\left(\frac{\pi}{4}\right) = \Delta\varphi_D(\frac{\pi}{2})$ is satisfied. In other words, the magenta curve representing $I_C I_D$ in Fig. 4(a) maintains the same minimum phase sensitivity as that of the black curve representing $I_D$.

**Conclusion**

We experimentally demonstrated, for the first time to the best of our knowledge, an enhanced resolution achieved via the N-fold intensity product of equally divided MZI output fields. This resolution enhancement was consistent with SNL feature in classical physics. Through analytical evaluation of the corresponding Fisher information, we confirmed that the phase sensitivity of the N-fold intensity product remains independent of N and does not degrade compared to the original (undivided) MZI output field. Furthermore, we demonstrated that the resolution of the cross-intensity product between two MZI output fields was enhanced by a factor of two without compromising the original phase sensitivity of each field. Compared to the *i.i.d.* N-subfield case, an additional $\sqrt{2}$ improvement in resolution arose from the intrinsic out-of-phase relation between the MZI outputs. A complete implementation of this scheme has recently been investigated under the framework of classically excited superresolution, which emulates the photonic de Broglie wavelength in quantum sensing. The *i.i.d.* data used for the uncorrelated N-fold intensity product was provided by Poisson distributed photons of a coherent laser input. When compared to the N00N-state-based quantum sensing, SNL-limited classical sensing with access to a vastly larger photon numbers can outperform Heisenberg-limited quantum schemes in practice, due to the current limitation on N of N00N states (N ≪ 100). Moreover, the enhanced resolution of the *i.i.d.* N-fold intensity product scales as $\sqrt{N}$, or $\sqrt{2N}$ when both MZI outputs are used. These results suggest that intensity-product sensing metrology may serve as a competitive alternative to conventional interferometer-based optical sensors, such as a fiber-optic gyroscopes and wavelength meters.

**Methods**

The Mach-Zehnder interferometer (MZI) shown in Fig. 1 consisted of two 50/50 non-polarizing beam splitters. A continuous-wave (CW) laser (Thorlabs HNL020L) with a wavelength of 633 nm and a coherence time (length) of 200 ns (60 m) was used as the light source. The incident polarization was aligned within the MZI plane. Four single-photon counting modules (SPCMs; Excelitas SPCM-AQRH-15), each with a dynamic range exceeding $3.5 \times 10^7$ photons/s, a dark count rate of approximately 50 photons/s, and a temporal resolving time of ~350 ps, were employed for N-fold intensity product measurements under a single-photon coincidence-detection scheme. For measuring intensity-product correlations at photon numbers N = 1 to 4, a four-channel coincidence counting unit (Altera, DE2) was used to post-select events from attenuated laser light. For the CW measurement results shown in Figs. 3(b) and 4(b), the input laser power prior to entering the MZI was set to be 3 µW ($\simeq 10^{13}$ photons/s). In this regime, fast silicon avalanche photodiodes (Thorlabs, APD-110A), each offering a linear detection response up to 1.5 µW, replaced the SPCMs. The coincidence counting unit was substituted with a four-channel digital oscilloscope (Yokogawa, DL9040), providing a maximum sampling rate of 5 GHz and a bandwidth of 500 MHz. Each data point in Fig. 3(b) represents an average over 30 samples. The N-fold intensity products from the APDs were post-processed using the recorded oscilloscope data. In both detection schemes used in Fig. 3, the piezoelectric phase shifter (Fig. 1) was scanned over a 36-second period in a forward direction from $\varphi = -2\pi$ to $\varphi = 2\pi$, with phase increments of $2\pi/90$, resulting in 180 data points per fringe curve in Fig. 4. The mean photon count per data point was collected over 0.1 seconds, yielding average values of $\langle R_D^{(1)} \rangle \sim 10^5$, $\langle R_D^{(2)} \rangle \sim 10^3$, and $\langle R_D^{(4)} \rangle \sim 10^1$.

**Data Availability**
All data generated or analyzed during this study are included in this published article.

**Author Contributions**

JS proposed experiments. SK conducted experiments. BSH supervised experiments, analyzed the data, and wrote the manuscript.

**Funding**

This work was supported by the IITP-ITRC grant (IITP 2025-RS-2021-II211810; 50%) funded by the Korean government (Ministry of Science and ICT). BSH also acknowledges that this work was supported by GIST via GIST research program 2025.

**Competing interests**

The authors declare no competing interests.